# FLOW-FORCE COMPENSATION IN A HYDRAULIC VALVE

**Jan Lugowski**
Purdue University
West Lafayette, IN, USA
lugowskj@purdue.edu

**ABSTRACT**

Flow-reaction forces acting in hydraulic valves have been studied for many decades. Despite this, they are difficult to account for due to the complexities of the jet flow. This paper focuses only on the reduction, also referred to as compensation, of the flow force as applied to a valve spool featuring a profile of a turbine bucket. Fluid power textbooks explain the compensation taking place on such a profile by applying Newton's laws of motion to the profile and deliver an equation for the magnitude and the direction of the flow force. This paper shows that both the magnitude and the direction of the compensating flow force are incorrect if calculated from the textbook equation. A corrected analysis of the dynamic forces is presented that are in agreement with earlier experiments by this author. It follows that the compensating flow force should be calculated from the static-pressure imbalance on the spool profile. That is, not Newton's but Pascal's law should be applied to calculate the compensating flow force.

**INTRODUCTION**

A hydraulic valve often controls a fluid flow through it by restricting its flow area. This restriction results in a jet downstream of the orifice and the accompanying force, known as flow induced force, hydraulic reaction force, or Bernoulli force [1]. The flow force is a design problem if it is acting on a moving part of the valve and is large enough to affect the operation of the valve. Pilot-operated valves, featuring large actuating force, have been a practical solution to the flow-force problem. But even for servo systems, it is important to know the axial forces necessary to operate the valve piston [2]. The need to understand the flow force is even greater if the valve is to be operated directly, without the pilot stage.

In their paper that first explained and accounted for the flow force, Lee and Blackburn [2] offered a design solution to reduce, or compensate, the flow force by shaping the valve piston, or spool, like a turbine bucket. They also provided a formula to compensate the magnitude of the compensated flow force. The formula and the theory behind it, based on Newton mechanics, has been later reprinted and expanded in textbooks, such as by Blackburn et al [3], Merritt [1], and Guillon [4].

To the best of this author's knowledge, no publication by other researchers can be found that has put the textbook theory in question. To be fair, the geometry of a turbine-bucket profile is more complex than that of a rocket engine or a water-turbine bucket. No literature could be found on how to calculate the flow reaction force on a turbine bucket of a valve spool. The profile on the spool consists of *two* adjacent turbine buckets. The fluid mechanics textbooks limit their analysis to only *one* turbine bucket.

This author attempted to compensate the flow force by shaping the spool as a turbine bucket and found experimentally that the flow force is reduced more if the entry angle of the profile is smaller (between 25° and 30°), much less than the suggested 69°. The experiments with various profile angles also revealed that the exit angle of the turbine-bucket profile did not contribute to the reduction of the flow force, contrary to the textbook suggestions. More experiments followed that showed that the compensating force acts on the spool profile *upstream* from the orifice, on the opposite end of its supposed location [5].

This paper explains how the flow force should be calculated by following Newton mechanics and attempts to clarify where and how errors have been made by the authors of the textbook theory. The errors are very difficult to recognize due to the simplicity of the momentum theory and the convincing explanation by its authors. Also, some textbooks mention the complexities of the jet flow and admit that the actual flow forces may differ from the calculated ones. It also does not help to see a problem when one deals with a well-established textbook theory. Still, the errors are big, rendering the theory useless and misleading. The actual forces on the turbine bucket would act in the *opposite* directions and have *quite different* magnitude.

**CONTROL VOLUME ANALYSIS OF JET FLOW**

In fluid mechanics, control volume (CV) is an arbitrary volume through which fluid flows. Control volume method is used to make the analysis of the flow more convenient, allowing focusing attention on a volume in space [6]. Figure 1 shows a jet impinging against a wall with velocity v. If the flow rate is Q and the fluid density is ρ, then the active force $F_y$ can be calculated as shown from Newton's laws of motion. The reaction force would have equal value but opposite direction. In this paper, only active forces will be considered, just as they are in the literature on flow force in hydraulic valves [1-4]. If the jet is perpendicular to the wall, there is no jet force acting parallel to the wall, if friction forces are disregarded.

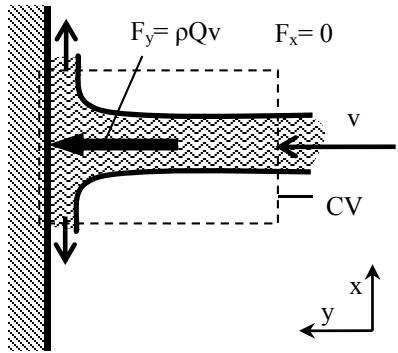

Fig. 1 Control volume applied to calculate jet force $F_y$

Even if the jet is not perpendicular to the wall, as in Fig. 2, there will be no active force $F_x$, parallel to the wall. This is due to the negligibly small exit velocity v of the jet out of the control volume CV.

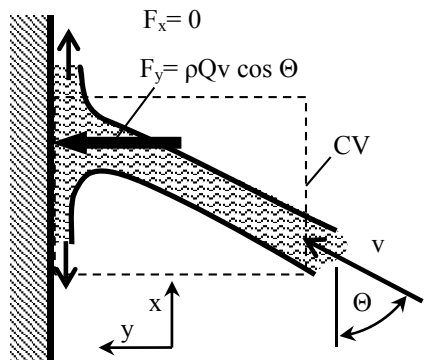

Fig. 2 Force $F_y$ from jet impinging on wall at angle Θ

When the jet is perpendicular to the wall and is deflected by 180° after impinging on the wall, while maintaining its velocity v, the force $F_y$ acting on the wall is twice as large, Fig. 3. There are two accompanying forces $F_x$ that cancel each other, acting to push the wall profile apart.

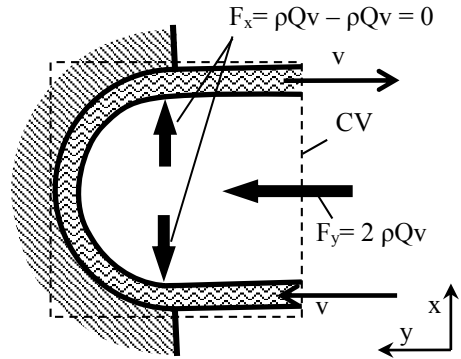

Fig. 3 Force $F_y$ from jet reflected from the wall

If we split the control volume into two adjacent ones $CV_1$ and $CV_2$, shown separated in Fig. 4 to help with the analysis of the forces, we can see why the total force $F_y$ is two times larger compared with Fig. 1. Force $F_{y1}$ is generated by the jet entering $CV_1$, and force $F_{y2}$ is generated by the jet exiting $CV_2$. Since both forces act pushing the wall to the left, the total force Fy is twice as large. The vertical force $F_{x1}$ is created by the jet exiting $CV_1$ and pushes the wall profile down. The other vertical force $F_{x2}$ is created by the jet entering $CV_2$ and pushes the wall profile up. The net vertical force $F_x$ is zero as both forces cancel out. Still, their effect is clear as they are pushing the wall profile apart. This is important to notice as the textbook theory ignored those forces. Ignoring them means that both vertical forces are zero and do not push the profile apart. This may not be important for many flow cases, but leads to a serious error if applied to a turbine-bucket profile on a spool in a hydraulic valve.

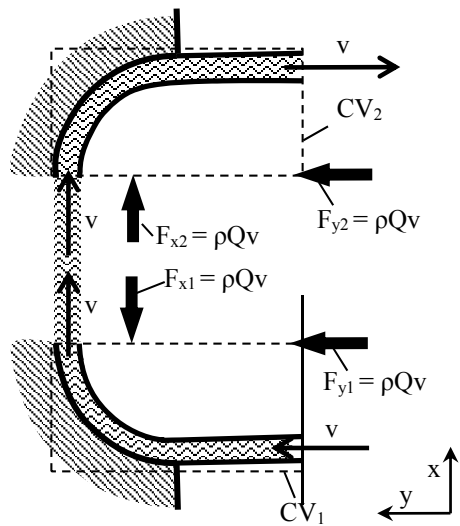

Fig. 4 Splitting control volume helps analyze forces by jet reflected from the wall

If the entry and exit angles are not 90°, as for a turbine-bucket profile, both the horizontal forces $F_{y1}$ and $F_{y2}$ will be smaller, as shown in Fig. 5. Again, $CV_1$ and $CV_2$ are adjacent to each other and shown apart only to visualize the forces. Now there are two vertical forces $F_x$ acting on each CV. On $CV_1$, force $F_{x11}$ is reduced because the jet angle is less than 90° and is produced by the jet entering $CV_1$. The opposite vertical force $F_{x12}$ is produced by the jest exiting $CV_1$ and has full strength because it is parallel to the vertical x axis. We assume that the jet maintains its velocity v along the profile, just like the textbook theory does. The net force $F_{x1}$ acting on profile within $CV_1$ is

$$F_{x1}= F_{x11} - F_{x12} = \rho Q v (\cos \Theta_1 - 1) \tag{1}$$

Force $F_{x1}$ acts vertically on the entry profile to push it down, as the force is negative. If we canceled both forces $F_{x12}$ and $F_{x21}$ because they act in the opposite directions, the net force $F_{x1}$ would be positive ($\rho Q v \cos \Theta_1$) and had a different value as well than that per Eq. (1).

Similarly, there are two vertical forces $F_x$ acting on $CV_2$, force $F_{x22}$ is reduced because the jet angle is less than 90° and is produced by the jet exiting $CV_2$. The opposite vertical force $F_{x21}$ is produced by the jest entering $CV_2$ and has full strength because it is parallel to the vertical x axis. The net force $F_{x2}$ acting on profile within $CV_2$ is

$$F_{x2}= F_{x21} - F_{x22} = \rho Q v (1 - \cos \Theta_2) \tag{2}$$

Force $F_{x2}$ acts vertically on the exit profile to push it up, as the force is always positive. Again, if we canceled both forces $F_{x12}$ and $F_{x21}$ because they act in the opposite directions, the net force $F_{x2}$ would be negative (- $\rho Q v \cos \Theta_2$) and have a different value as well than that per Eq. (2).

The axes in the figures have been rotated by 90° on purpose to help visualize all the forces acting on a wall, and how to properly apply Newton's laws of motion to calculate them. The coordinate system will be rotated back when discussing a spool profile in a hydraulic valve.

Notice that the *two horizontal forces* $F_y$ add up while the vertical forces $F_x$ subtract from each other. There are *four vertical forces* $F_x$ and two of them are of identical magnitude but acting in opposite directions. Those two forces, $F_{x12}$ and $F_{x21}$, are only visible when we split the control volume into $CV_1$ and $CV_2$. Fluid power and fluid mechanics textbooks do not discuss a turbine-bucket profile that is bent more than 90°. If they did, they would need to apply two control volumes to properly account for all the vertical forces. Failing to do so would result in calculating forces acting in the opposite directions, and of wrong magnitude.

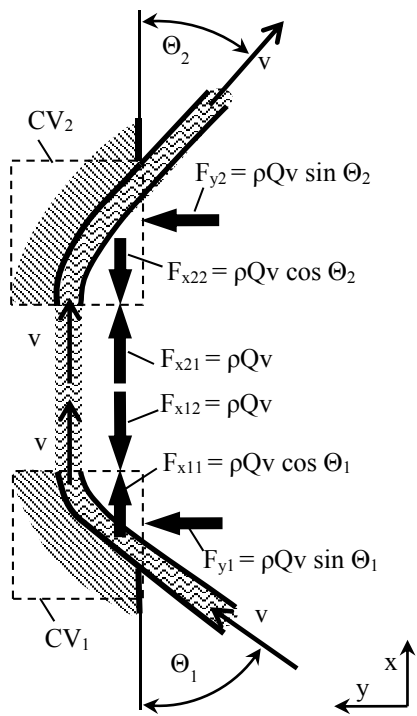

Fig. 5 Forces by jet reflected from a turbine-bucket profile

Both vertical forces $F_{x1}$ and $F_{x2}$ are shown in Fig. 6 depicting the turbine-bucket profile, with jet entering it at angle $\Theta_1$ and exiting it at angle $\Theta_2$. The net vertical force $F_x$ acting on the turbine-bucket profile is

$$F_x = F_{x1} + F_{x2} = \rho Q v \, [(\cos \Theta_1 - 1) + (1 - \cos \Theta_2)] \tag{3}$$

Now it is easier to see why one would be tempted to cancel both vertical forces $F_{x12}$ and $F_{x21}$: Equation (3) could be simplified without changing the magnitude and sign of the net axial force $F_x$:

$$F_x = F_{x1} + F_{x2} = \rho Q v \, (\cos \Theta_1 - \cos \Theta_2) \tag{4}$$

At first look, this approach seems reasonable. But doing this has serious consequences if Eq. (4) were applied to select entry and exit angles $\Theta_1$ and $\Theta_2$ of the turbine-bucket profile. Despite this, the textbooks provide Eq. (4) as correct. And, following the incorrect Eq. (4), they advise to design the profile featuring a large entry angle $\Theta_1$ to reduce the positive force acting upwards to close the valve. But the actual force $F_{x1}$ is negative and acts downwards on the profile. Per Eq. (1) a larger angle $\Theta_1$ produces a larger, not smaller, force $F_{x1}$. The bigger problem, though, is that the correct force acts downwards, not upwards.

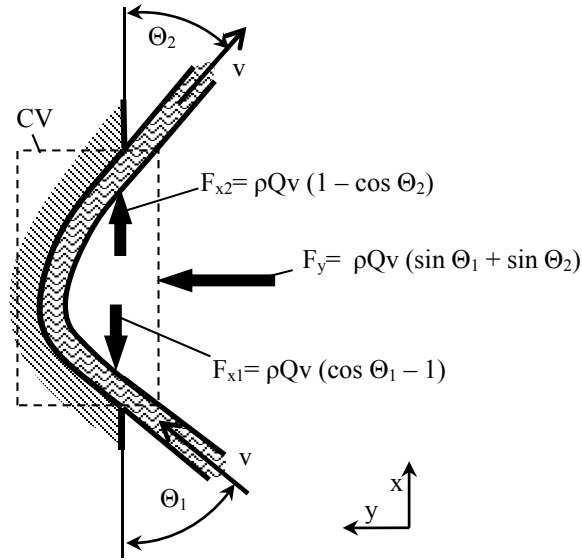

Fig. 6 Forces by jet reflected from a turbine-bucket profile

Figure 7 shows the same turbine-bucket profile from Fig. 6 but oriented horizontally, as it is customary in the fluid power literature. The x-axis is parallel to the spool axis. The radial force $F_y$ cancels itself on the spool circumference and has no effect on the axial force $F_x$.

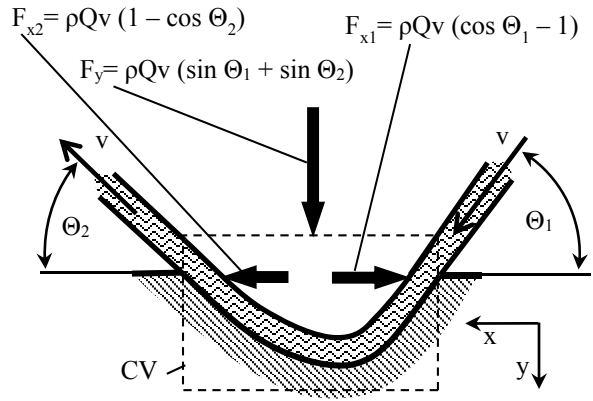

Fig. 7 Forces on a spool featuring a turbine-bucket profile are pushing the profile apart

Figure 8 shows only axial forces $F_x$ with the incorrect textbook formula, see also Eq. 4, to calculate them. It is worth mentioning that the textbook formula indicates that the closing, positive, flow force $F_{x1}$ is produced on the entry profile while the opening, negative, flow force $F_{x2}$ is produced on the exit profile. This advice is not correct and leads to confusion if applied to design a spool profile. The biggest problem with the textbook theory is that it ignores the spreading of the jet due to turbulence and the accompanying loss of jet velocity v. For this reason exit profile cannot produce the compensating flow force due to the small jet velocity. Even if the jet did maintain its velocity, this exit profile would produce *a positive* flow force, not a compensating one. While it is true that the turbine-bucket profile compensates the flow force, the compensation does not take place on the exit profile located downstream from the valve orifice but on a very small area of the entry profile located *upstream* from the valve orifice [5]. This is simply due to the uncompensated high static pressure pushing on the spool to open the valve. Thus, the textbook formula per Eq. (4) is even more misleading as the negative compensating force should be calculated by applying Pascal's, not Newton's, law.

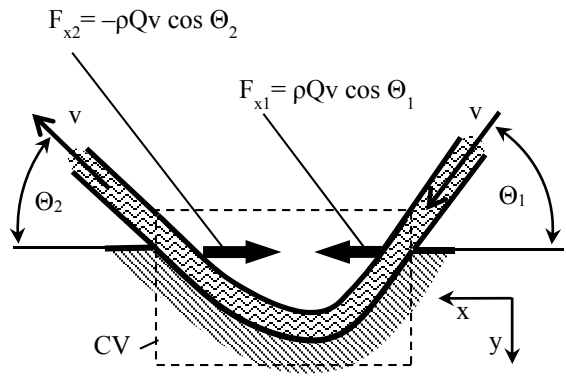

Fig. 8 Forces on a spool featuring a turbine-bucket profile per fluid power textbooks [1-4]

**SUMMARY**

The analysis of the forces acting on the spool shaped like a turbine bucket points to where the authors of the textbook theory made errors in their attempt to provide a formula for calculating the flow force on such a profile. The first error is due to canceling of the two opposite forces which are invisible if one only considers one control volume. This resulted in wrong directions of the flow forces, and wrong magnitudes. It helps little that the sum of the two wrong forces is the same as the sum of the correct *four* forces. As the exit profile produces *positive*, not the compensating, flow force, the textbook theory wrongly puts a value on the exit profile to deliver the compensating force.

The presented analysis agrees with earlier experimental data by this author. The compensating force is produced on the entry profile *upstream* from the orifice. The authors of the textbook theory made another mistake by disregarding the static-pressure balance on the spool profile and focused only on the dynamic forces. Since the pressure at the inlet of the valve can be very high, even a small area on the spool profile can produce a significant *compensating* flow force,

**NOMENCLATURE**

| | |
|---|---|
| $\rho$ | Mass density of fluid, kg m$^{-3}$ |
| $\Theta$, | Jet angle, deg. |
| $\Theta_1$ | Entry jet angle, deg. |
| $\Theta_2$ | Exit jet angle, deg. |
| F | Flow force, N |
| $F_x$ | Axial flow force, N |
| $F_{x1}$ | Axial flow force to close the valve (positive), N |
| $F_{x2}$ | Axial flow force to open the valve (negative, compensating), N |
| $F_y$ | Radial flow force, N |
| Q | Flow rate, m$^3$s$^{-1}$ |
| v | Velocity of jet at vena contracta, m s$^{-1}$ |

**ACKNOWLEDGMENTS**
The author is grateful for support by Purdue University's Maha Fluid Power Educational Advisory Board and by Parker Hannifin Corporation, Hydraulic Valve Division.